\documentclass[10pt]{iopart}
\newcommand{\BF}[1]{\mbox{\boldmath$#1$}}    
\def\la{\label}
\def\al{\alpha} 
\def\bet{\beta} 
\def\bea{\begin{eqnarray}}
\def\eea{\end{eqnarray}}
\def\bsea{\begin{subeqnarray}}
\def\esea{\end{subeqnarray}} 
\def\be{\begin{equation}}
\def\ee{\end{equation}}

\begin{document}

\title[Glass Model with Mode Coupling and
  Trivial Hamiltonian]
{Dynamic Mean-Field Glass Model with Reversible Mode Coupling and
  Trivial Hamiltonian}

\author{Kyozi Kawasaki\dag\footnote[3]{Permanent address: 4-37-9 Takamidai, Higashi-ku, Fukuoka
  811-0215, Japan}  and Bongsoo Kim \ddag  
} 
\address{\dag\ CNLS, Los Alamos National Laboratory,
Los Alamos, NM 87545, USA}

\address{\ddag\ Physics Dept., Changwon National University, 
Changwon, 641-773, Korea}

\begin{abstract}
Often the current mode coupling theory(MCT) of glass transitions is
compared with mean field theories. We explore this possible
correspondence. After showing a simple-minded derivation of MCT with
some difficulties we give a concise account of our toy model developed
to gain more insight into MCT. We then reduce this toy model by
adiabatically eliminating rapidly varying velocity-like variables to
obtain a Fokker-Planck equation for the slowly varying density-like
variables where diffusion matrix can be {\it singular}. This gives a
room for non-ergodic stationary solutions of the above equation.  
\end{abstract}




\section{Introduction}
\setcounter{equation}{0}
Our understanding of phase transition starts from the famous thesis of
van der Waals of 1873 where complex effects of intermolecular
interactions were put into two parameters often denoted as
$b$ and $a$ representing, respectively, the repulsive and attractive
parts of intermolecular forces. This theory was transcribed into
magnetism resulting in the Weiss theory of ferromagnetism which
contains a single parameter measuring strength of the molecular
field. These mean field theories were remarkably successful and provided a good
beginning of phase transition theory. Shortcomings of these theories
were soon noticed, especially after Onsager's exact analysis of the
two-dimensional Ising model. Thus the basis of the mean field theories
were examined and exactly solvable model was constructed which
yielded mean field theory results \cite{kuh61}. The characteristics of this model is that
the attractive part of intermolecular potential has an infinite range
after a certain limiting procedure, which suppress fluctuation effects
responsible for deviations from the mean field behavior. Efforts to
incorporate neglected fluctuation effects led to our current day
understanding of phase transition, in particular,  critical phenomena.

In comparison, the case of structual glass is much less clear-cut \cite{jj86}. First there
are still uncertainties about nature of the glass transition. It is
not clear whether there is a real transition or just a
cross-over. Eventhough one assumes a genuine transition, opinions
differ among those who think that there is an underlying thermodynamic
transition and those who believe that the transition is merely a
kinetic one. The case of spin glass is much more well understood, at
least on the mean field level where we know the existence of a genuine
thermodynamic phase transition \cite{mpv87}.  

Under such circumstances it is quite significant that the first
principle theory of structural glass transition, that is, the MCT,
was proposed and succeeded in explaining some aspects of glass transitions\cite{bgs84,wg91}.
Since the theory is still rather crude, it is fair to regard the status of
 this theory as similar to that of
van der Waals and Weiss theories mentioned above.  As in the case of
these theories the current MCT is beset with serious
difficulties. These basically come from the fact that the MCT
formalism was originally developed for critical phenomena focussing on
very large length scales reaching to thousands of \AA
ngstroms. Applications to glasses have difficulties due to
its primarily short length scales and due to its own peculiarities. These difficulties are:
(a) The  factorization approximation which replaces
the four-body time corrrelation functions by the product of two-body
time correlation functions is essential to obtain the self-consistent MCT
equation. This is especially uncontrolled at short length scales of at
most 10-20 \AA ngstroms. See the following section. 
(b) The idealized MCT predicts a sharp dynamic transition to a nonergodic state at
a certain temperature. But MCT does not provide any information on the nature of this 
nonergodic state.
(c) The physical picture of the so called hopping processes in an extended version of MCT is still lacking.

It should then be an urgent task for further progress to clarify the bases of the MCT.
Motivated by this desire we constructed a toy model
having the following  three features \cite{kkbk}:
\begin{itemize}
\item reversible mode coupling mechanism
\item trivial statics
\item mean-field type so that the model can be exactly solvable.
\end{itemize}
This toy model is distinguished from other toy models for glasses in
that it closely mimics the MCT \cite{bgs84,wg91}. 


\section{MCT}
\setcounter{equation}{0}
Here we present an over-simplified derivation of a self-consistent MCT
equation of the density-density time correlation function obtained
first in \cite{bgs84,wg91}. 
We start from the following hydrodynamics-like continuum equation \cite{dm}:
\bea
m\frac{\partial}{\partial t}\rho({\bf r},t)=-{\BF \nabla}\cdot {\bf
  j(r},t)
\la{eqn:2-1a}\\
\frac{\partial}{\partial t} {\bf j(r},t)= {\bf f(r},t) +\cdots
\la{eqn:2-1b}
\eea
Here $m$ is the mass of a fluid molecule, $\rho({\bf r},t)$ is the
number density, and ${\bf j(r},t)$ is the momentum density. The
ellipsis in
 (\ref{eqn:2-1b}) contains the terms second order in the
momentum density, dissipative terms and thermal noise terms, which do
not play a role here and will be dropped hereafter.

Here we assume the existence of a free energy density functional
$H(\{\rho \})$. Then the body force density ${\bf f(r},t)$
 is the number density
times a force on a test particle of the same kind. The latter is
negative of the gradient of the infinitesimal variation of the free energy density
functional against infinitesimal density change, that is, $-{\BF
  \nabla}\delta H(\{\rho \})/\delta \rho({\bf r})$. Therefore we find
(Here time arguments are omitted.)
\be
 {\bf f(r)}=-\rho({\bf r}){\BF \nabla}\delta H(\{\rho \})/\delta \rho({\bf r})
\la{eqn:2-2}
\ee

Now, the exact form for $H(\{\rho \})$ is unknown and various
approximate forms are proposed. For the purpose of deriving the MCT
equation, it suffices to use the popular Ramakrishnan-Yussouf form
despite its shortcomings \cite{ry79}:
\bea
H(\{\rho \})=k_BT\int d{\bf r}\rho({\bf r})\left[ \ln\Big(\frac{\rho({\bf
    r})}{\rho_0}\Big)-1 \right] \nonumber \\  -\frac{1}{2}k_BT \int d{\bf r}\int d{\bf
  r}'c(|{\bf r-r}'|)\big(\rho({\bf r})-\rho_0 \big)\big(\rho({\bf
  r}')-\rho_0 \big)
\la{eqn:2-3}
\eea
Here $\rho_0$ is the density of the reference uniform liquid and
$c(|{\bf r-r}'|)$ is the direct correlation function \cite{hm86} of the reference liquid. 
The Fourier transform $\hat c({\bf k})$ of $c(r)$ is connected
with the static structure factor of reference liquid $S({\bf k})$ through
\be
\hat c({\bf k})=\rho_0^{-1}-S({\bf k})^{-1}
\la{eqn:2-4}
\ee
The fact that the direct correlation function appears here is
important because this is the  only place in this theory where the short range
correlation central to any liquid theory is incorporated.  

The next step is to split ${\bf f(r},t)$ into terms linear and
quadratic in the density difference $\delta \rho({\bf r},t)\equiv
\rho({\bf r},t)-\rho_0$ as
\bea
{\bf f(r},t)={\bf f}^l({\bf r},t)+{\bf f}^{nl}({\bf r},t) 
\la{eqn:2-5a} \\
{\bf f}^l({\bf r},t)=-{\BF \nabla}p({\bf r},t)
\la{eqn:2-5b}  \\
{\bf f}^{nl}({\bf r},t)=k_BT\int d{\bf r}'c(|{\bf r-r}'|)\delta
\rho({\bf r},t){\BF \nabla}'\delta \rho({\bf r}',t)
\la{eqn:2-5c}
\eea
where $p({\bf r},t)$ is the local pressure correct up to $\delta
\rho({\bf r,t})$. Combination of (\ref{eqn:2-1a}),(\ref{eqn:2-1b}) and
(\ref{eqn:2-5a}),(\ref{eqn:2-5b}) and (\ref{eqn:2-5c}) tells us that the ${\bf f}^l$ produces only linear
density oscillations with constant wave vectors and does not
contribute to freezing. On the other hand, ${\bf f}^{nl}({\bf r},t)$ is
the sum of numerous terms oscillating with different frequencies,
which, on the whole, look quite irregular.  This fact also makes it
hopeless to try to find solutions to these equations. 

However, we are not interested in individual
solutions but only some statistical properties of them, which are also
measurable quantities. Among such quantities the most attention is paid to 
the density-density time correlation function, whose normalized form
is defined in terms of $\rho_{\bf k}(t)$, the Fourier transform of $\delta\rho({\bf r},t)$, by
\be
\phi_k(t)\equiv \frac{<\rho_{\bf k}(t)\rho_{-\bf k}(0)>}{<\rho_{\bf
    k}(0)\rho_{-\bf k}(0)>}
\la{eqn:2-6}
\ee
In obtaining this quantity from (\ref{eqn:2-1a}) and (\ref{eqn:2-1b})
we can regard  ${\bf f}^{nl}({\bf r},t)$ as a kind of random force
familiar in the Langevin equation of Brownian motion \cite{nk92}. Then the
equation that determines $\phi_k(t)$ requires knowledge of a memory
kernel which is the time correlation function of ${\bf f}_{\bf
  k}^{nl}(t)$, the Fourier transform of ${\bf f}^{nl}({\bf r},t)$.

Explicitly the equation for $\phi_k(t)$ turns out to be \cite{kk00}
\be
\frac{d^2\phi_k(t)}{dt^2}=-\Omega_k^2\phi_k(t)-\int_0^tds{\cal
  M}_k(t-s)\frac{d\phi_k(s)}{ds}
\la{eqn:2-6a}
\ee
where $\Omega_k\equiv k\sqrt{k_BT/S_k}$ is the frequency of the local
density oscillation in liquids. ${\cal
  M}_k(t)$ is the memory kernel given by
\be
{\cal M}_k(t)=\frac{1}{m\rho_0k_BTk^2V}\big<f_{\bf k}^{nl}(t)f_{-\bf
  k}^{nl}(0)\big>
\la{eqn:2-6b}
\ee
where $f_{\bf k}^{nl}\equiv i{\bf k \cdot f_k}^{nl}$ and $V$
the system volume.
Since this correlation function involves products of  four density  
fluctuations which are impossible to deal with directly, 
this is factorized into products of two density-density
corrrelation functions. This produces the feedback mechanism
responsible for freezing. The resulting self-consistent equation for the
density-density correlation function is the same as  that given for
the first time in\cite{bgs84}.
In the above simple derivation we have side-stepped the fact that the
time dependence of  ${\bf f}^{nl}({\bf r},t)$ in the memory kernel is
in fact governed by the ``projected'' dynamics in the sense of
Zwanzig-Mori formalism. The correct but more involved derivation
starts from a Fokker-Planck type equation \cite{kk00}. Alternative
simpler derivation explicitly relies on the strong  assumption that density
fluctuations at various times obey Gaussian statistics \cite{zfstd01}. 

The transition to non-ergodic states in this theory is driven by the
nonlinear force term (\ref{eqn:2-5c}), which, in turn arises from the
quadratic (or harmonic) term of (\ref{eqn:2-3}). The first term there
containing a logarithm is just for ideal gas. Therefore, the
Hamiltonian of this theory does not require complex nonlinear terms
that characterize many other theories or models of glass transitions.


\section{Mean field toy model}
\subsection{Model}
Our toy model is a set of oscillators with linear and random nonlinear
couplings expressed by the following Langevin equations for the 
$N$-component density-like variables $a_i(t)$ with $i=1,2,\cdots,N$ and 
the $M$-component velocity-like variables $b_{\al}$ with $\al=i,2, \cdots, M$.
Here and after we will use Roman indices for the components of $a$ and 
Greek for those of $b$:
\bea
\dot{a}_i= K_{i\al}b_{\al}+\frac{\omega}{\sqrt{N}} J_{ij\al}a_j b_{\al} \la{EQ1} \\ 
\dot{b}_{\al}  = -\gamma b_{\al}-\omega^2 K_{j\al}a_j   
-\frac{\omega}{\sqrt{N}}J_{ij\al} (\omega^2a_ia_j- T\delta_{ij})+f_{\al}  \la{EQ2} \\
 <f_{\al}(t)>=0, \qquad <f_{\al}(t)f_{\bet}(t')> = 2\gamma T\delta_{\al \bet} \delta (t-t') \la{EQ3}
\eea
where the summation is implied for repeated indices and overdots
denote time derivatives.
Here $\gamma$ is the decay rate of the velocity-like variables
$b_{\al}$ and $\omega$ gives a measure of the  frequencies of
oscillations of the density-like variables $a_j$. 
The thermal noises $f_{\al}(t)$  are independent Gaussian random variables with zero mean
 and variance $2\gamma T$, $T$ being the temperature of the heat bath 
with which the system has a thermal contact.
The choice of this variance guarantees the proper equilibration of the
variables $\{b\}$.
The $N \times M$ matrix $K_{i\al}$ plays an important role in the model and 
for later purpose we impose the (one-sided) orthogonality
\be
K_{i\al}K_{i\bet}=\delta_{\al\bet}, \qquad K_{i\al}K_{j\al}\neq \delta_{ij}
\la{EQ4}
\ee
where the last equation is due to the inequality $M<N$.
For $M=N$ we can impose an additional condition $K_{i\al}=\delta_{i\al}$ and hence trivially
 $K_{i\al}K_{j\al}=\delta_{ij}$.
We also note that  $K_{i\al}$ governs linearized reversible dynamics of the model 
with the dynamical matrix $\bf \Omega$ given by $\Omega_{ij} \equiv \omega^2K_{i\al}K_{j\al}$. 
The mode coupling coefficients $J_{ij\al}$ are chosen to be quenched (time-independent) Gaussian 
random variables with the following properties:
\bea
\overline{J_{ij\alpha}}^J =0,  \nonumber \\
\overline{J_{ij\alpha}J_{kl\beta}}^J = \frac{g^2}{N}
\biggl[(\delta_{ik}\delta_{jl}+\delta_{il}\delta_{jk})\delta_{\alpha
\beta} + K_{i\beta}(K_{k\al}\delta_{jl}+K_{l\al}\delta_{jk}) \nonumber \\  
+ K_{j\beta}(K_{k\al}\delta_{il}+K_{l\al}\delta_{ik})\biggr] \la{EQ5}
\eea
where $\overline{\cdots }^J$ denotes average over the $J$'s.
In constructing this model, we were motivated by the works \cite{kraich,bouch} in which
random coupling models involving an infinite component order parameter have been shown 
to be exactly analyzed by mean-field-type concepts.

Equation (\ref{EQ1}) is analogous to the equation of continuity of fluid and 
 (\ref{EQ2}) is like the equation of motion where the right hand side is like
the force acting on a fluid element which corresponds to
(\ref{eqn:2-1a}) and (\ref{eqn:2-1b}), respectively.
We will eventually take $N$ and $M$ infinite with the ratio
$\delta^* \equiv M/N$ kept finite.

One can derive from the Langevin equations (1)-(3) the corresponding Fokker-Planck equation 
for the probability distribution function $D(\{a\},\{b\},t)$ for
our variable set denoted as  $\{a\},\{b\}$  as follows
\be
\partial_t D(\{a\},\{b\},t) =  {\hat L} D(\{a\},\{b\},t) \la{EQ6}
\ee
where the Fokker-Planck operator is given by 
${\hat L}={\hat L}_0 + {\hat L}_1+ {\hat L}_{MC}$ \\
with  
\be
\eqalign{{\hat L}_0 \equiv  \frac{\partial}{\partial b_{\alpha}} 
\gamma \left(T \frac{\partial}{\partial
b_{\alpha}}+b_{\alpha} \right), \quad
 {\hat L}_1 \equiv K_{j \al} \left(- \frac{\partial}{\partial a_j}
b_{\al}+\frac{\partial}{\partial b_{\al}} \omega^2
a_j \right), \nonumber \\
 {\hat L}_{MC}  \equiv \frac{1}{\sqrt{N}} J_{ij\alpha}\left(
-\frac{\partial}{\partial a_i}\omega a_jb_{\alpha} +
\frac{\partial}{\partial b_{\alpha}} \omega (\omega^2
a_ia_j-T\delta_{ij})\right)} \la{EQ7}
\ee
It is then easy to show that the {\it equilibrium} stationary distribution 
(i.e., ${\hat L}D_e({a},{b})=0$) is given by
\be
 { D}_e(\{a\},\{b\}) = cst. \rme^{-\sum_{j=1}^N\frac{\omega^2}{2T}a_j^2-
\sum_{\alpha=1}^M\frac{1}{2T}b_{\alpha}^2} \la{EQ8}
\ee
where $cst.$ is the normalization factor.

\subsection{Analysis  and discussion}
We aim at finding the set of five equilibrium time correlation functions defined by
\be
\eqalign{\fl C_a(t-t') \equiv  \frac{1}{N} <a_j(t)a_j(t')>,  \quad
 C_{ab}(t-t') \equiv \frac{1}{M}K_{j\al}<a_j(t)b_{\al}(t')>,  \\
 \fl C_{ba}(t-t') \equiv \frac{1}{M}K_{j \al}<b_{\al}(t)a_j(t')>,  \quad
  C_b(t-t') \equiv \frac{1}{M} <b_{\alpha}(t)b_{\alpha}(t')>,  \\
  C_a^K(t-t') \equiv \frac{1}{M}K_{i\al}K_{j\al}<a_i(t)a_j(t')>} \la{EQ14}
\ee
It turns out that we need to have the last correlation function to close the self-consistent
 set of equations for the correlators when $M<N$.
Note that for the case $M=N$, we can take $K_{i\al}=\delta_{i\al}$, and
then $C_a^K(t-t')=C_a(t-t')$.

In order to obtain this self-consistent set of equations,  it is most convenient to adapt the generating
functional method from which one can write down the set of effective linear Langevin equations
valid in the limit of $M,N \rightarrow \infty$. We refer \cite{kik01} for further details.
>From this effective Langevin equations, one can readily derive the
following closed self-consistent equations for $t>0$
for the {\em five} correlators: 
\bea
\dot C_a(t) = \delta^*C_{ba}(t)
-\Sigma_{aa}\otimes C_a(t)-\delta^*\Sigma_{ab}\otimes C_{ba}(t) \la{EQ26} \\
\dot C_{ba}(t) = -\gamma C_{ba}(t)-\omega^2 C_a^K(t)
-\Sigma_{ba}\otimes C_a^K(t)-\Sigma_{bb}\otimes C_{ba}(t) \la{EQ27}\\
\dot C_{ab}(t) = C_b(t)
-\Sigma_{aa}\otimes C_{ab}(t)-\Sigma_{ab}\otimes C_b(t), \la{EQ28} \\
\dot C_b(t) = -\gamma C_b(t)-\omega^2 C_{ab}(t)
-\Sigma_{ba}\otimes C_{ab}(t)-\Sigma_{bb}\otimes C_b(t) \la{EQ29}\\
\dot C_a^K(t) = C_{ba}(t)
-\Sigma_{aa}\otimes C_a^K(t)-\Sigma_{ab}\otimes C_{ba}(t) \la{EQ30}
\eea
where, for any function $X(t)$, $X \otimes a(t) \equiv \int_{-\infty}^t dt' \,
X(t-t')a(t')$.
The equations (\ref{EQ26})-(\ref{EQ30}) constitute the self-consistent equations for the 5 correlators
$C_a(t)$, $C_{ba}(t)$,  $C_{ab}(t)$, $C_b(t)$, and $C_a^K(t)$.
This set of equations can be solved numerically with the initial conditions
$C_a(0)=C_a^K(0)=T/\omega^2$, $C_{ab}(0)=C_{ba}(0)=0$, and $C_b(0)=T$.
Here the kernels $\Sigma$'s are given by
\be
\eqalign{\fl \Sigma_{aa}(t-t') \equiv \delta^*\frac{g^2\omega^4}{T}\bigl(C_a(t-t')C_b(t-t')+
\delta^* C_{ab}(t-t')C_{ba}(t-t')\bigr), \\
\Sigma_{ab}(t-t') \equiv -2 \delta^* \frac{g^2\omega^4}{T}C_a(t-t')C_{ba}(t-t')  \\
\Sigma_{ba}(t-t') \equiv -2 \delta^* \frac{g^2\omega^6}{T}C_a(t-t')C_{ab}(t-t'),  \\
 \Sigma_{bb}(t-t')\equiv \frac{2g^2\omega^6}{T}C_{a}(t-t')^2} \la{EQ22}
\ee  
These kernels arise from the non-linear mode coupling terms in
(\ref{EQ6}) and (\ref{EQ7}).
Note that the correlator $C_a^K(t,t')$ is not involved in the $\Sigma$'s.

For further analyses it is very convenient to work with the equations of
the Laplace transformed correlation functions defined as
$C^L(z) \equiv \int_0^{\infty} dt\, \rme^{-zt} \, C(t)$.
Performing the Laplace transformation of the self-consistent equations
and rearranging them we obtain
\bea
\fl C^L_a(z)=\frac{T}{\omega^2}\frac{1}{z+\Sigma^L_{aa}(z)}
\biggl[1-\delta^* \frac{\omega^2(1-\Sigma^L_{ab}(z))^2}{ 
(z+\Sigma^L_{aa}(z))(z+\gamma+\Sigma_{bb}(z))+ \omega^2(1-\Sigma^L_{ab}(z))^2} \biggr] \la{EQ36} \\
C^L_{ab}(z)=-\frac{T(1-\Sigma^L_{ab}(z))}{(z+\Sigma^L_{aa}(z))
(z+\gamma+\Sigma_{bb}(z)) +\omega^2(1-\Sigma^L_{ab}(z))^2} \la{EQ37}\\
C^L_{ba}(z)=\frac{T(1-\Sigma^L_{ab}(z))}{(z+\Sigma^L_{aa}(z))
(z+\gamma+\Sigma_{bb}(z)) +\omega^2(1-\Sigma^L_{ab}(z))^2}\la{EQ38} \\
C^L_b(z)=\frac{T(z+\Sigma^L_{aa}(z))}{(z+\Sigma^L_{aa}(z))
(z+\gamma+\Sigma_{bb}(z)) +\omega^2(1-\Sigma^L_{ab}(z))^2}\la{EQ39}\\
C^{KL}_a(z)=\frac{T}{\omega^2}\biggl[z+\Sigma^L_{aa}(z)
+\frac{\omega^2(1-\Sigma^L_{ab}(z))^2} {z+\gamma+\Sigma_{bb}(z)} \biggr]^{-1} \la{EQ40} 
\eea


For $\delta^*=1$ where $M=N$ and $K_{i\al}=\delta_{i\al}$, 
$C^L_a(z)=C^{KL}_a(z)$ reproduces the equation derived in \cite{sdd}, apart from
the wave number dependence.
Note that if we put $\Sigma^L_{aa}(z)=\Sigma^L_{ab}(z)=0$ {\it by hand},
(\ref{EQ36}) or (\ref{EQ40}) gives a closed equation for $C_a(t)$ alone. This equation is nothing but
the Leutheusser's schematic MC equation giving a dynamic transition from
ergodic phase to nonergodic one \cite{bgs84}. But in reality $\Sigma_{aa}$ and 
$\Sigma_{ab}$ can not be ignored {\it a priori} and our numerical solution strongly indicates
that  the system remains ergodic for all temperatures due to the 
strong contribution of these so called hopping terms. 
Furthermore these hopping terms cannot be made self-consistently small 
as temperature is lowered. Therefore the 
density correlator does not show a continuous slowing down with lowering temperature.
This result was   striking to us  since usually a mean-field-type theory, such as
the dynamics of the spherical $p$-spin model in the limit of $N \rightarrow \infty$, 
often gives a sharp dynamic transition \cite{cris}. 

Thus it is very difficult to understand the idealized MCT {\em without} relying upon
uncontrolled approximations. 
It is also interesting to note that the ergodicity restoring process
in our toy model (represented by the kernels $\Sigma_{aa}$ and $\Sigma_{ab}$) 
has nothing to do with a thermally activated energy barrier crossing since the quadratic 
Hamiltonian in our model does not possess such a barrier.


\section{Reduced Fokker-Planck equation for the density-like variables}
\la{sec:fp} 
Possibility of nonergodic states in our model can be seen more directly by 
adiabatically eliminating the variables $\{b\}$ in the
limit of large $\gamma$ and obtaining the reduced Fokker-Planck equation
for the distribution function $\tilde D (\{a\},t)$
containing only the $\{a\}$ variables:
\bea
\frac{\partial \tilde D(\{a\},t)}{\partial t} ={\cal L}_{FP} \tilde
D(\{a\},t)\nonumber \\\equiv
 \frac{\partial}{\partial a_i}
 Q_{ij}(\{a\}) \left( \frac{\partial }{\partial a_j}
+\frac{\omega^2} {T}a_j \right) \tilde D(\{a\},t)  
\la{eqn:fp-1}
\eea
where we  have defined the Fokker-Planck operator ${\cal L}_{FP}$ through the
second member of the above equation.
Here the diffusion matrix $Q_{ij}(\{a\})$ is given by
\bea
Q_{ij}(\{a\})\equiv \frac{T}{\gamma} M_{i\al} M_{j\al} 
\la{eqn:fp-2a} \\
M_{i\al} \equiv K_{i\al}+ \frac{\omega}{\sqrt{N}}J_{ik\al}a_k
\la{eqn:fp-2b}
\eea
The positive semi-definiteness of the diffusion tensor ${\bf Q}$ is very easy
to show because for an arbitrary $N$ component real vector we have
\be
x_iQ_{ij}x_j=\frac{T}{\gamma}\sum_{\al}\big(M_{i\al}x_i\big)^2\geq 0
\la{eqn:fp-4}
\ee

The crucial point is that the diffusion matrix $Q_{ij}$ is {\em
  singular} for $M<N$, i.e., 
det$|{\bf Q}|=0$  giving rise to zero eigenvalues for $\bf
  Q$ \cite{kik01}. 
This implies that the Fokker-Planck equation (\ref{eqn:fp-1})
can have {\em nonequilibrium} stationary solution other than 
the equilibrium one, $\tilde D_e(\{a\})=cst. \exp (-\omega^2 a_j^2/2T)$.
This nonequilibrium stationary solutions are precisely the kind of 
nonergodic states found numerically in the present toy model.
In fact a class of stationary solutions is given by
\be
\tilde D_s(\{a\})={\cal F}(\xi_j a_j)\,\rme^{-\frac{\omega^2}{2T}a_i^2}
\la{eqn:fp-3}
\ee
where $\xi_i$ is an eigenvector of the diffusion matrix $Q_{ij}$ with
zero eigenvalue and ${\cal F}(x)$ a non-negative function.
If the function ${\cal F}(x)$ is a constant, then
$\tilde D_s(\{a\})=\tilde D_e(\{a\})$ is the equilibrium distribution, otherwise it is a nonequilibrium
stationary distribution. Hence the model is nonergodic for $0 \leq
\delta^* < 1$.

Let us now investigate this point somewhat further.
We first define the following non-negative ratio:
\be
R(\{a\},t) \equiv \frac{\tilde D(\{a\},t)}{\tilde D_e(\{a\})}
\la{eqn:fp-5}
\ee
and then introduce a Boltzmann's H-like quantity as 
\be
{\cal H}(t)\equiv \int d\{a\}\tilde D(\{a\},t)\ln R(\{a\},t)
\la{eqn:fp-6}
\ee
where the integration is over all the variables in the set $\{a\}$.
Using normalization property of the distribution function and 
integrating by parts assuming a natural boundary condition we find
\be
\dot{\cal H}(t)=\int d\{a\}\tilde D(\{a\},t){\cal 
L}_{FP}^{\dagger}(\{a\})\ln R(\{a\},t)
\la{eqn:fp-6a}
\ee
with ${\cal L}_{FP}^{\dagger}(\{a\})$ the adjoint operator of ${\cal L}_{FP}(\{a\})$:
\be
{\cal L}_{FP}^{\dagger}(\{a\})\equiv \Big(\frac{\partial}{\partial
  a_i}-\frac{\omega^2}{T}a_i\Big)Q_{ij}(\{a\})\frac{\partial}{\partial
  a_j}
\la{eqn:fp-7}
\ee
Now we readily verify that
\be
{\cal L}_{FP}^{\dagger}(\{a\}) \ln R(\{a\},t) = \frac{1}{R}{\cal 
L}_{FP}^{\dagger}R-Q_{ij}\frac{1}{R}\frac{\partial R}{\partial a_j}\frac{1}{R}\frac{\partial 
R}{\partial a_i}
\la{eqn:fp-8}
\la{eqn:4.11}
\ee
Substituting this into (\ref{eqn:fp-6a})  we finally find
\be
\dot{\cal H}(t)=-\int d\{a\} \tilde D(\{a\},t)Q_{ij} (\{a\})\frac{\partial \ln R(\{a\},t)}{\partial 
a_i}\frac{\partial \ln R(\{a\},t)}{\partial a_j}
\la{eqn:fp-9}
\ee
 where the contribution of the first term in (\ref{eqn:fp-8}) vanishes since
\be
\fl \int d\{a\} \frac{\tilde D}{R} {\cal L}_{FP}^{\dagger}\cdots =
\int d\{a\} \tilde D_e {\cal L}_{FP}^{\dagger}\cdots =
\int d\{a\} \left( {\cal L}_{FP}\tilde D_e \right) \cdots =0
\la{eqn:fp-10}
\ee
This result shows a kind of Boltzmann's H-theorem \cite{risken} (See \cite{nk92} for
a discussion for general master equations with detailed balance.):
\be
\dot{\cal H}(t) \leq 0
\la{eqn:fp-11}
\ee
Let us suppose that we have performed a transformation of the
variables $\{a\}\rightarrow \{s\},\,\,\tilde D(\{a\})d\{a\}\rightarrow
\hat D(\{s\})d\{s\},\,\,R(\{a\})\rightarrow \hat R(\{s\})$ etc. so that the matrix $Q_{ij}(\{a\})
$ is diagonalized:
\be
\hat Q_{ij}(\{s\})=\lambda_i(\{s\})\delta_{ij}
\la{eqn:fp-12}
\ee
with the eignevalues $\lambda_i(\{s\})$  non-negative functions of $\{s\}$.
For the case of diagonalized $Q_{ij}$, (\ref{eqn:fp-11}), the results
(\ref{eqn:fp-9}) and (\ref{eqn:fp-10}) become
\be
\int d\{s\}\sum_i\lambda_i(\{s\})\hat D(\{s\},t)\Bigg(\frac{\partial
  \ln \hat R(\{s\},t)}{\partial s_i}\Bigg)
^2 \geq 0
\la{eqn:fp-13}
\ee
The stationarity condition $\frac{\partial}{\partial t}\hat D(\{s\},t)=0$
 or $\dot{ \cal H}(t)=0$ then implies 
\be
\int d\{s\}\sum_i\lambda_i(\{s\})\hat D(\{s\},t)\Bigg(\frac{\partial
  \ln \hat R(\{s\},t)}{\partial s_i}\Bigg)
^2 = 0
\la{eqn:fp-14}
\ee
If we denote those subset of the variables of $\{s\}$ with positive eigenvalues as
$s_{\al}, \,\, \al =1,2,\cdots , M'(\leq M)$, which we denote as
$\{s\}'$,
we must have in the region $\hat D(\{s\},t)>0$,
\be
\frac{\partial \ln \hat R(\{s\},t)}{\partial s_{\al}} = 0
\,\,\mbox{for}\,\,\lambda_{\al}(\{s\})>0
\la{eqn:fp-15}
\la{eqn:4.17}
\ee
We note that validity of the condition (\ref{eqn:fp-14}) in general
depends on the regions in the space of the variables
$\{s\}$ through $\{s\}$ dependence of the $\lambda$'s. Thus, in a
particular region in which   $\lambda_{\al}(\{s\})\hat D(\{s\})$ are
positive definite, $ R(\{s\},t)$ does not contain $s_{\al}$. If we
denote the remaining set of the variables $\{s\}$ than those
corresponding to positive eigenvalues as $u_{\bet},\,\,
\bet=M'+1,M'+2, \cdots, N$, which are corresponding to the zero eigenvalues, 
we should have $\hat R(\{s\},t)=\hat R(\{s\}',\{u\},t)=\hat R(\{u\},t)$. 
Consequently, from (\ref{eqn:fp-5}), 
the general form of non-equilibrium stationary state distribution function
is 
\be
\hat D_S(\{s\}',\{u\})=\hat R_S(\{u\})\hat D_e(\{s\}',\{u\})
\la{eqn:fp-16}
\la{eqn:4.18}
\ee
The previous result, (\ref{eqn:fp-3}), is a  special case of this general form.

\section{Summary and discussion}
Our toy model analyses show that the so-called hopping processes
introduced as a correction to the idealized MCT \cite{wg91} are the
effects of the velocity-like variables, which still enter in the
framework of the meanfield treatment, and are not directly related to
the barrier-crossing mechanism. In our toy model, strength of the
hopping processes can be adjusted through the parameter $\delta^*$.
Possible nonequilibrium stationary states are connected to the
singularity of the diffusion matrix of reduced Fokker-Planck equation
of Section \ref{sec:fp} involving only the varaibles $\{a\}$. This
feature, suitably extended to general master equation, would be
common to many kinetically constrained glass transition models 
 \cite{bcn} and is worth further exploration.

\ack
We thank K. Dawson, M. Fisher, W. G\"{o}tze,  J. J\"{a}ckle,  A. Latz, S. J. Lee,
M. M\'ezard, R. Schilling and C. Tsallis for useful suggestions and discussions.
The research of KK is supported by the Department of Energy, under contract
W-7405-ENG-36. An additional partial support to KK by the Cooperative
Research under the Japan-U.S. Cooperative Science Program sponsored by
Japan Society of Promotion of Science is also gratefully aknowledged.
BK is supported by the Interdisciplinary Research Program of the KOSEF (Grant No.
1999-2-114-007-3).

\end{document}